\documentclass[10pt,twocolumn,letterpaper]{article}

\usepackage{cvpr}              

\usepackage{graphicx}
\usepackage{amsmath}
\usepackage{amssymb}
\usepackage{booktabs}

\usepackage{times}
\usepackage{epsfig}
\usepackage{multirow}
\usepackage[dvipsnames]{xcolor}
\usepackage{caption}
\usepackage{subcaption}
\usepackage{bm}
\usepackage{python}
\usepackage[frozencache,cachedir=.]{minted}

\usepackage[accsupp]{axessibility}

\definecolor{medgreen}{RGB}{0, 200, 0}

\usepackage[pagebackref,breaklinks,colorlinks]{hyperref}

\usepackage[capitalize]{cleveref}
\crefname{section}{Sec.}{Secs.}
\Crefname{section}{Section}{Sections}
\Crefname{table}{Table}{Tables}
\crefname{table}{Tab.}{Tabs.}

\def\cvprPaperID{3} 
\def\confName{CVPR}
\def\confYear{2023}

\begin{document}

\title{QuickSRNet: Plain Single-Image Super-Resolution Architecture \\for Faster Inference on Mobile Platforms}

\author{Guillaume Berger\thanks{Contributed equally.} \hspace{0.2em} Manik Dhingra\footnote[1]{} \hspace{0.2em} Antoine Mercier \hspace{0.2em} Yashesh Savani \hspace{0.2em} Sunny Panchal \hspace{0.2em} Fatih Porikli \\
Qualcomm AI Research\thanks{Qualcomm AI Research is an initiative of Qualcomm Technologies, Inc.}\\
{\tt\small \{guilberg, manidhin, amercier, ysavani, sunnpanc, fporikli\}@qti.qualcomm.com}
}

\maketitle
\thispagestyle{empty}

\begin{abstract}
    
    In this work, we present QuickSRNet, an efficient super-resolution architecture for real-time applications on mobile platforms. Super-resolution clarifies, sharpens, and upscales an image to higher resolution. Applications such as gaming and video playback along with the ever-improving display capabilities of TVs, smartphones, and VR headsets are driving the need for efficient upscaling solutions.  While existing deep learning-based super-resolution approaches achieve impressive results in terms of visual quality, enabling real-time DL-based super-resolution on mobile devices with compute, thermal, and power constraints is challenging. To address these challenges, we propose QuickSRNet, a simple yet effective architecture that provides better accuracy-to-latency trade-offs than existing neural architectures for single-image super-resolution. 
    We present training tricks to speed up existing residual-based super-resolution architectures while maintaining robustness to quantization. Our proposed architecture produces 1080p outputs via 2$\times$ upscaling in 2.2 ms on a modern smartphone, making it ideal for high-fps real-time applications.



\end{abstract}


\section{Introduction}

\begin{figure*}[th!]
  \centering
  \begin{subfigure}{0.23\linewidth}
    \centering
    \includegraphics[width=\linewidth]{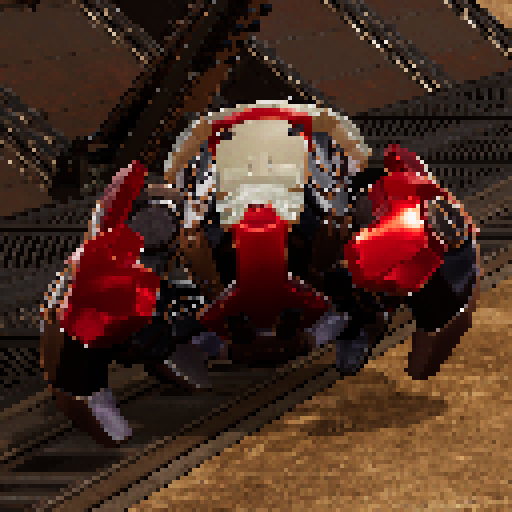}
    \caption{LR}
  \end{subfigure}
  \hspace{1mm}
  \begin{subfigure}{0.23\linewidth}
    \centering
    \includegraphics[width=\linewidth]{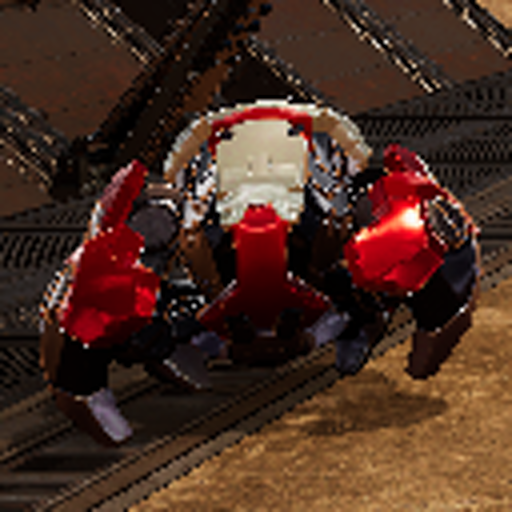}
    \caption{Interpolation}
  \end{subfigure}
  \hspace{1mm}
  \begin{subfigure}{0.23\linewidth}
    \centering
    \includegraphics[width=\linewidth]{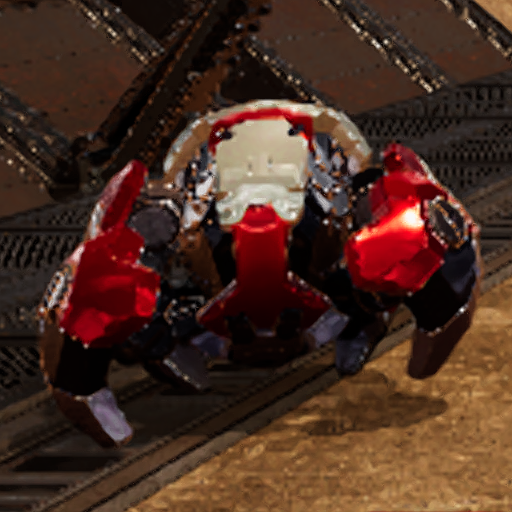}
    \caption{Algorithmic approach}
  \end{subfigure}
  \hspace{1mm}
  \begin{subfigure}{0.23\linewidth}
    \centering
    \includegraphics[width=\linewidth]{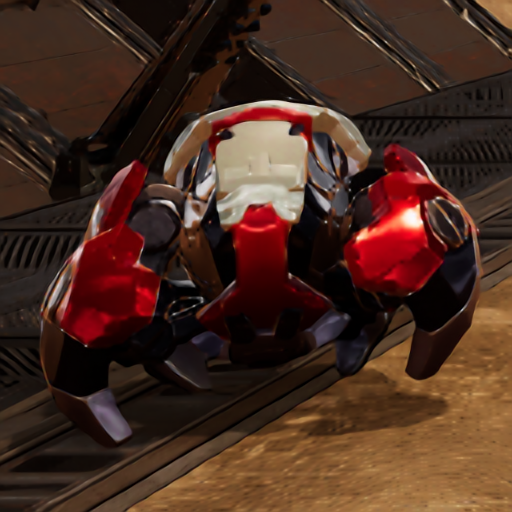}
    \caption{ML-based approach}
  \end{subfigure}
  \caption{Side-by-side comparison of various upscaling approaches: (a) Low-resolution, (b) Bicubic interpolation, (c) Non-ML baseline: FSR1.0 \cite{fsr}, (d) ML baseline: EDSR \cite{edsr}.}
  \label{fig:bicubic-fsr-edsr}
\end{figure*}

Single-image super-resolution (SR) refers to a family of techniques that recover a high-resolution (HR) image $I_{HR}$ from its low-resolution (LR) counterpart $I_{LR}$. In recent years, deep learning (DL) based approaches have become increasingly popular in the field \cite{srcnn, fsrcnn, espcn, imdn, hbpn, edsr, rfdn, hat, sr3}, producing impressive results compared to interpolation-based techniques and hand-engineered heuristics (see \cref{fig:bicubic-fsr-edsr}). However, most existing DL-based super-resolution solutions are computationally intensive and not suitable for real-time applications requiring interactive frame rates, such as mobile gaming. While DL-based super-resolution has been successfully applied to gaming on high-end GPU desktops \cite{dlss, xess}, neural approaches are still impractical for mobile gaming due to their high latency and computational costs. For example, a DL-based architecture such as EDSR \cite{edsr} takes $75$ ms to upscale a $540$p image to $1080$p on a state-of-the-art mobile AI accelerator. This has driven the need for efficient DL-based super-resolution solutions \cite{xlsr, abpn, sesr, repsr, rendersr} that can be used in real-time applications such as video gaming, where responsiveness and higher frame rates are essential.


\begin{figure}[t]
    \centering
    \includegraphics[width=\linewidth]{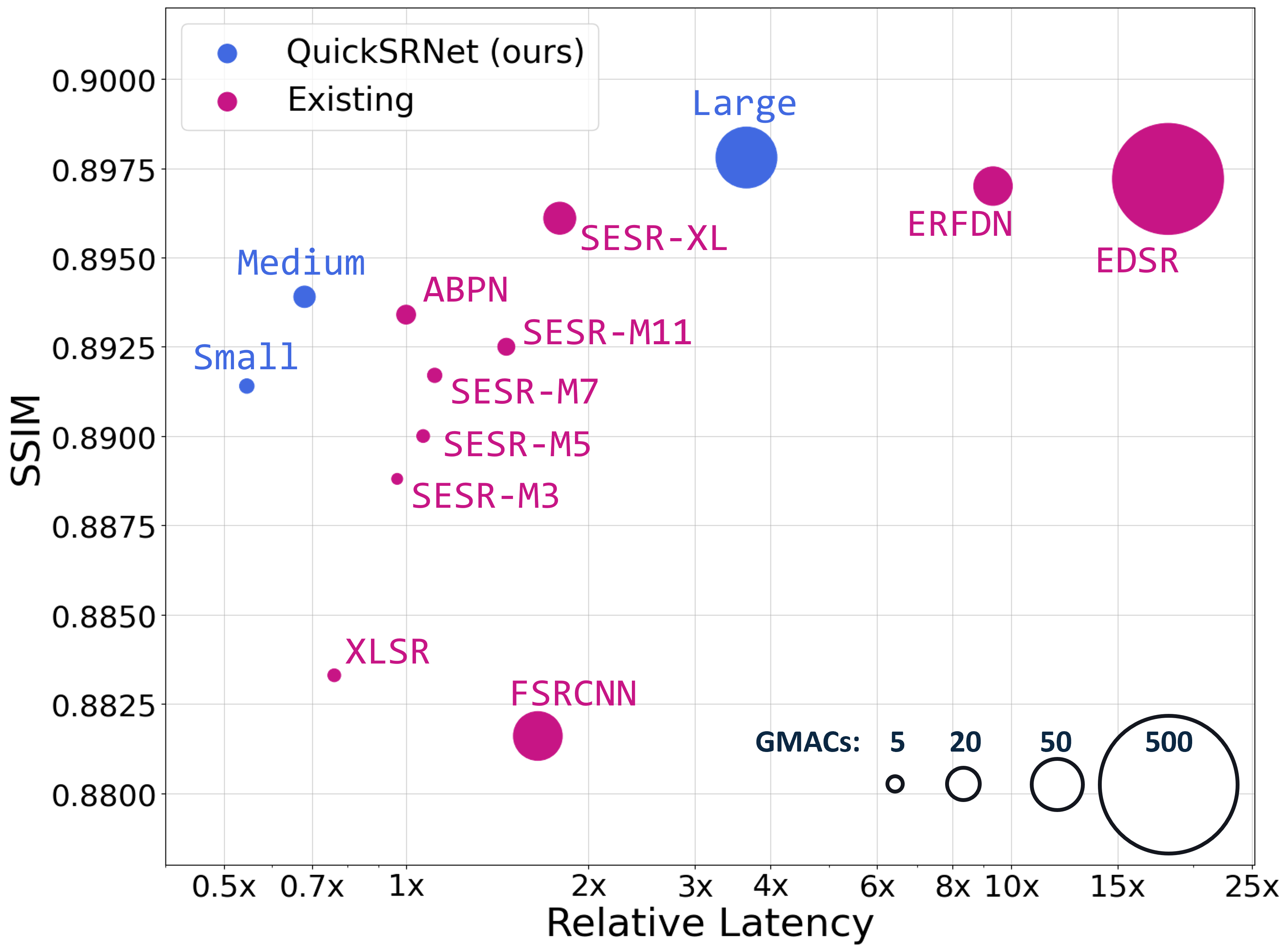}
    \caption{Accuracy-to-latency trade-offs of QuickSRNet (ours) against existing SISR architectures. We report accuracy after 8-bit quantization and measure latency on a state-of-the-art mobile AI accelerator.}
    \label{fig:quicksrnet-perf}
\end{figure}

In this work, rather than trying to achieve the state-of-the-art PSNR or SSIM scores on standard super-resolution benchmarks, we aim to develop efficient architectures that are suitable for high-fps real-time applications on mobile devices. To this end, we propose QuickSRNet, a simple single-image super-resolution neural network that obtains better accuracy-to-latency trade-offs than existing efficient SR architectures.  In particular, we make the following key contributions:

\begin{itemize}
\item We streamline the network architecture, reduce the impact of residual connection removal, and ultimately demonstrate the effectiveness of simpler designs in achieving high levels of accuracy and on-device performance.
\item We compare a wide variety of architectures in terms of on-device latency, measured on a device with Snapdragon\textsuperscript{\textregistered} 8 Gen 1 Mobile Platform, instead of FLOPS count, which is not a reliable indicator of on-device performance \cite{heim2021measuring}. 
\item We measure accuracy \emph{after} 8-bit quantization, a necessary step for better efficiency on mobile platforms, and describe architectural tricks that improve robustness to quantization. 
\item We apply our proposed architecture to a real-world use-case (video gaming) and compare its visual quality against that of a well-known industrial non-ML based approach (AMD's FidelityFX Super Resolution (FSR1.0) algorithm \cite{fsr}). 
\item We describe an approach to perform $1.5\times$ upscaling, a setting that is occasionally used in gaming and XR use-cases but not trivially supported by SR architectures whose upscaling step is based on a sub-pixel convolution \cite{espcn}.
\end{itemize} 

\begin{figure*}[th!]
\begin{center}
\includegraphics[width=\textwidth]{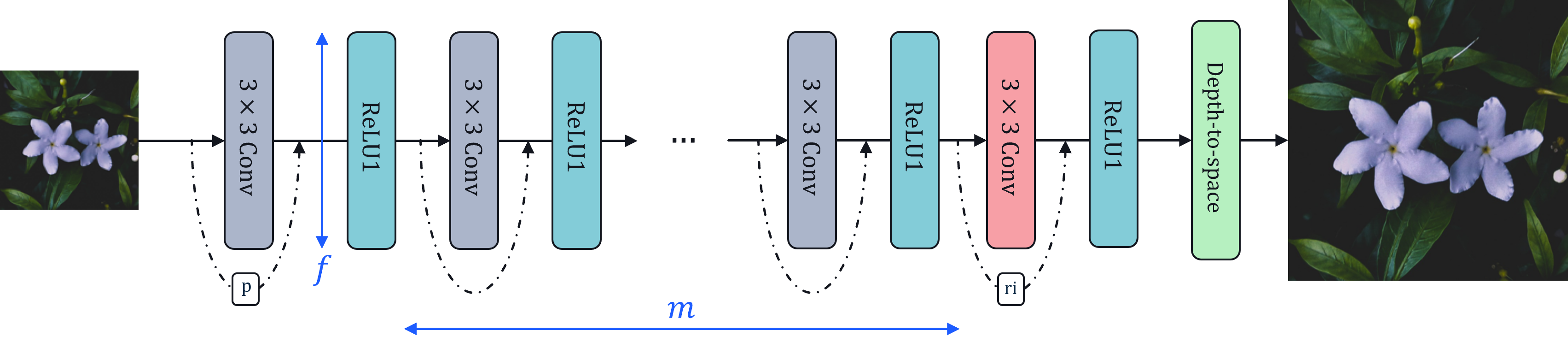}
\end{center}
  \caption{QuickSRNet architecture. We use the convention \textit{QuickSRNet-f$X$-m$Y$} to refer to the architecture variant that has $Y$ intermediate conv layers and $X$ feature channels. We use dotted lines to illustrate that the conv layers are initialized using an identity initialization scheme. In practice, these skip connections are incorporated into the weights of the corresponding conv module. $p$ and $ri$ stand for ``partial" and ``repeat-interleaving" respectively (see \cref{subsec:quicksrnet} for more details).}
\label{fig:qsrnet-base-architecture}
\end{figure*}

\section{Related work}

Several efficient SR architectures have been proposed recently. Overall, these architectures share many characteristics with the earlier work by \cite{fsrcnn} and \cite{espcn} on FSRCNN and ESPCN respectively: they are usually fully convolutional, use a relatively small number of layers and channels, all layers run at the input resolution and the final output is mapped to higher resolution using a subpixel convolution\footnote{In the rest of the paper, we will use the term \textit{depth-to-space} operation. In practice, a subpixel convolution amounts to performing a regular convolution producing $3\times S^{2}$ low-resolution channels, where $S$ is the scaling factor, followed by a \textit{depth-to-space} operation to map to higher resolution.}. Compared to these baselines, more recent approaches have incorporated the following changes:

\paragraph{XLSR} \cite{xlsr} uses grouped convolutions to reduce the computational footprint of the architecture and ``clipped" ReLU activations to improve robustness to quantization.

\paragraph{ABPN} \cite{abpn} employs a VGG-like convnet \cite{vgg} (i.e. consisting of only $3\times3$ Conv-ReLU blocks) with an ``anchor-based" input-to-output residual connection. This ``anchor-based" connection adds a channel-wise nearest-neighbor upscaled version of the input to the output before the final depth-to-space operation. We confirmed that this channel-wise implementation runs faster on our profiling device than the more common approach of adding the spatially-upsampled input directly to the output. Thus, we follow the same strategy to implement input-to-output residual connections in all our experiments. 

\paragraph{SESR} \cite{sesr} leverages linear over-parameterized residual modules which are collapsed into regular convolutions during inference for improved on-device performance. Other modifications include the use of long residual connections.

\paragraph{RepSR} \cite{repsr} investigates training VGG-like super-resolution architectures. Like ABPN, their convnet is equipped with a nearest-neighbor upsampling-based input-to-output connection. Similar to SESR, they find that using over-parameterized networks during training can boost accuracy. They propose a training scheme for using Batch Normalization (BN) layers \cite{batchnorm} without introducing artifacts in flat regions of the image, a typical side effect of BN when employed for super-resolution. At test time, the over-parameterized, BN-equipped network is collapsed into a simpler, more efficient network. 

\paragraph{Residual learning for super-resolution} Many SR architectures utilize a long skip connection which adds an upscaled version of the input $U(I_{LR})$ directly to the output. Efficient architectures (like \cite{abpn, repsr}) will often implement $U$ as nearest-neighbour interpolation. During training, SR architectures equipped with this technique are implicitly optimized to produce a residual $R = I_{HR} - U(I_{LR})$. One benefit is that the network produces reasonable outputs right after initialization which stabilizes training. Additionally, the input-to-output connection makes the architecture significantly more robust to quantization, as discussed in the next section.

\section{Methodology}

This section contains a detailed description of QuickSRNet as well as implementation details. The process for developing our proposed SR architecture began with preliminary experiments, which we present in the next paragraph.  

\subsection{On the impact of removing the input-to-output residual connection}

\begin{table}[t!]
  \setlength{\tabcolsep}{0.3em}
  \centering
  \small
  \begin{tabular}{|l|cc|c|}
    \hline
    \multicolumn{1}{|c|}{\multirow{2}{*}{\textbf{\begin{tabular}[c]{@{}c@{}}Architecture\end{tabular}}}} & \multicolumn{2}{c|}{\textbf{PSNR (dB)}} & \multirow{2}{*}{\textbf{Latency (ms)}} \\ \cline{2-3}
    \multicolumn{1}{|c|}{} & \multicolumn{1}{c|}{\textbf{FP16}}  & \textbf{INT8} &  
    \\ \hline
    ABPN & \multicolumn{1}{c|}{31.84 \footnotesize{(baseline)}} & 31.80 \footnotesize{(baseline)} & 2.17 \footnotesize{(baseline)}
    \\ \hline
    Res.-free ABPN & \multicolumn{1}{c|}{31.75 \textcolor{red}{($-$0.09)}}            & 31.50 \textcolor{red}{($-$0.30)} & 1.42 \textcolor{medgreen}{($-$35\%)} 
    \\ \hline
    \begin{tabular}[c]{@{}l@{}}QSRNet-Med\end{tabular} & \multicolumn{1}{c|}{31.82 \textcolor{gray}{($-$0.02)}} & 31.77 \textcolor{gray}{($-$0.03)} & 1.42 \color{medgreen}{($-$35\%)} 
    \\ \hline
  \end{tabular}
  \caption{On the accuracy and latency impacts of removing the input-to-output residual connection from the ABPN architecture. We report PSNR numbers obtained on BSD100 via $2\times$ upscaling. Latency numbers were obtained on a device with Snapdragon 8 Gen 1 using an input resolution of $512\times512$.}
  \label{tab:preliminary-experiments}
\end{table}

VGG-style architectures such as ABPN \cite{abpn} or RepSR \cite{repsr} are already well-optimized, so it is unclear how much faster they can be made on mobile AI accelerators. Instinctively, reducing the number of layers and channels, or replacing $3\times3$ kernels with $1\times1$ kernels, can improve speed at the cost of accuracy. Instead, our experiments investigate how to effectively remove the input-to-output residual connection without affecting accuracy.

As observed by \cite{xlsr, abpn}, long residual connections can have a large impact on the efficiency of super-resolution architectures, particularly on memory-limited platforms such as smartphones or VR headsets. 
To confirm this, we trained and profiled a residual-free ABPN variant and found that removing the input-to-output residual connection improves latency by $35\%$. However, this modification resulted in a marginally lower accuracy and more importantly, reduced robustness to quantization, as can be seen in \cref{tab:preliminary-experiments}. A similar trend is evident in the results of the Mobile AI 2022 Challenge \cite{ignatov2023efficient}, where the fastest approaches do not use input-to-output residual connections at the cost of accuracy. To address this, we propose QuickSRNet, a residual-free architecture which is robust to quantization.


\subsection{QuickSRNet}
\label{subsec:quicksrnet}

\begin{figure*}[t!]
    \captionsetup[subfigure]{justification=centering}
    \begin{minipage}{\textwidth}%
        \begin{minipage}{0.35\textwidth}%
            \begin{subfigure}{\linewidth}%
                \centering
                \includegraphics[height=5cm]{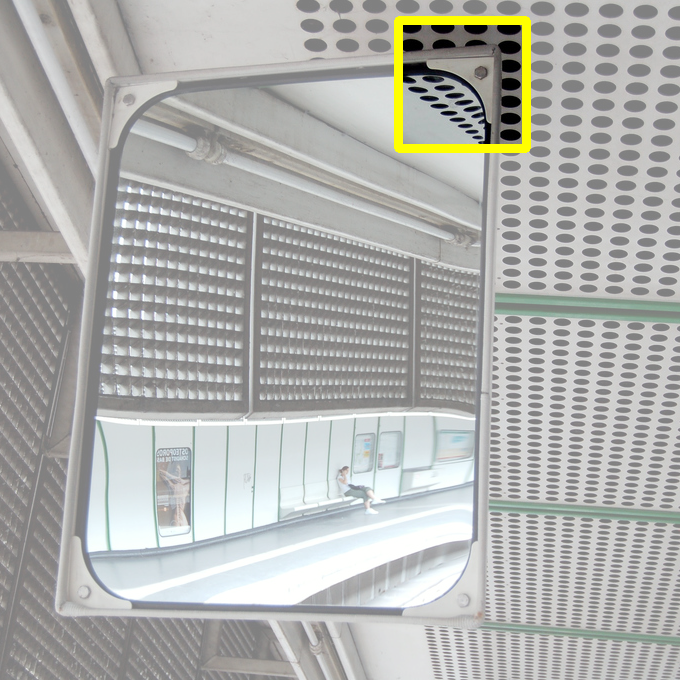}
                \vspace{-1mm}
                \caption*{}
                \label{fig:merge-scenario}
            \end{subfigure}
        \end{minipage}
        \begin{minipage}{0.8\textwidth}%
          \begin{subfigure}{0.2\linewidth}
            \centering
            \includegraphics[height=2.6cm]{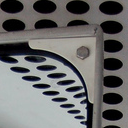}
            \vspace{-1mm}
            \caption*{HR}
          \end{subfigure}
          \begin{subfigure}{0.2\linewidth}
            \centering
            \includegraphics[height=2.6cm]{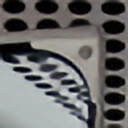}
            \vspace{-1mm}
            \caption*{SESR-M7}
          \end{subfigure}
          \begin{subfigure}{0.2\linewidth}
            \centering
            \includegraphics[height=2.6cm]{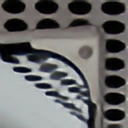}
            \vspace{-1mm}
            \caption*{ABPN}
          \end{subfigure}
          \begin{subfigure}{0.2\linewidth}
            \centering
            \includegraphics[height=2.6cm]{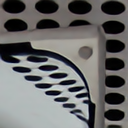}
            \vspace{-1mm}
            \caption*{EDSR}
          \end{subfigure}
          
          \vspace{1mm}
          
          \begin{subfigure}{0.2\linewidth}
            \centering
            \includegraphics[height=2.6cm]{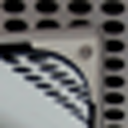}
            \vspace{-1mm}
            \caption*{Bicubic\\\phantom{empty}}
          \end{subfigure}
          \begin{subfigure}{0.2\linewidth}
            \centering
            \includegraphics[height=2.6cm]{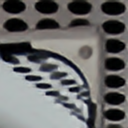}
            \vspace{-1mm}
            \caption*{Ours-Small\\(1.5$\times$ faster)}
          \end{subfigure}
          \begin{subfigure}{0.2\linewidth}
            \centering
            \includegraphics[height=2.6cm]{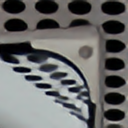}
            \vspace{-1mm}
            \caption*{Ours-Medium\\(1.3$\times$ faster)}
          \end{subfigure}
          \begin{subfigure}{0.2\linewidth}
            \centering
            \includegraphics[height=2.6cm]{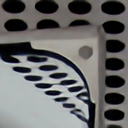}
            \vspace{-1mm}
            \caption*{Ours-Large\\(1.8$\times$ faster)}
          \end{subfigure}
        \end{minipage}
    \end{minipage}
    \caption{Visualization of 4× super-resolved images from Urban100 produced by our models and existing baselines. Our models match the quality of existing architectures while being significantly faster.}
    \label{fig:sr-comp-one-col}
    
\end{figure*}

Our architecture, QuickSRNet, follows a VGG-like structure with no input-to-output residual connection (see \cref{fig:qsrnet-base-architecture}). This architecture is denoted by $m$, the number of intermediate convolutional blocks, and $f$, the number of feature channels in those intermediate layers. To increase robustness to quantization, we use a residual learning-motivated initialization scheme along with clipped ReLU activations:

\paragraph{Identity initialization} We utilize an intuitive initialization technique where each intermediate convolutional layer simulates a localized skip connection:

\begin{equation}
    y = W \circledast x + x
    \label{eq:id-init}
\end{equation}

where $\circledast$ is the discrete convolution operator and $W$ refers to the kernel weights. In practice, we collapse the skip connection into the conv module: $y = (W + I) \circledast x = \hat{W} \circledast x$, where $\hat{W}$ are the modified weights after collapse and $I$ is the identity of discrete convolution operators. In this case, collapsing amounts to adding a diagonal of ones to the randomly initialized kernel sliced at the spatial center: $\forall i, W[i, i, c_x, c_y] \mathrel{+}= 1$  (with $c_x=c_y=1$ assuming a $3\times3$ kernel). This approach is akin to identity initialization \cite{idinithardt2016, idinitzhang2019fixup, idinitbachlechner2021rezero, idinitzeroinit} and related to the over-parameterized networks used in \cite{diracnet, repvgg, sesr, repsr}, except we collapse before training, during initialization. 

\Cref{eq:id-init} only works if $x$ and $y$ have the same dimensions and is therefore not directly applicable to the first and last layer of the architecture, as these layers respectively change the number of channels from $3$ to $f$ and $f$ to $3 \times S^{2}$, where $S$ is the scaling factor. For these layers, we modify the initialization scheme as follows:

\begin{itemize}
    \item \textbf{Partial identity initialization}: the $3$-channel input to the first convolutional module are added to the first $3$ output channels and the other $f - 3$ output channels are left unchanged.
    
    \begin{equation}
        y_i = 
        \begin{cases}
            (W \circledast x)_i + x_i,& \text{if } 0 \leq i < 3\\
            (W \circledast x)_i,              & \text{otherwise}
        \end{cases}
    \label{eq:id-init-partial}
    \end{equation}
    
    \item \textbf{Repeat-interleaving identity initialization}: the first $3$ input channels to the final convolutional module are repeat-interleaved $S^2$ times and added to the output.
    
    \begin{equation}
     y_i = (W \circledast x)_i + x_{round \left( \frac{i}{s^2}  \right)}
    \label{eq:id-init-repeat-interleave}
    \end{equation}
\end{itemize}

Similar to \cref{eq:id-init}, the skip connections described in \cref{eq:id-init-partial,eq:id-init-repeat-interleave} are incorporated into the corresponding convolutional module by adding ones to the kernel weights at the appropriate location. Intuitively, this initialization technique makes the input image propagate well throughout the entire network. The repeat-interleaving scheme used to initialize the final layer mimics the nearest-neighbour upscaling typically performed in the input-to-output connection of existing residual architectures. 

\paragraph{ReLU1} In addition to identity initialization, we found that clipping ReLU activations between $0$ and $1$ improves robustness to quantization. Compared to XLSR \cite{xlsr}, we use ReLU1s throughout the entire network as opposed to just the final layer. Note that for this approach to work well with our id-initialized architecture, it is important to scale input pixels between $0$ and $1$ (centering around $0$ would cause roughly half the pixels propagated by the first id-initialized conv to be zeroed out).

\paragraph{} Our experimental results (\cref{sec:experiments}) show that combining identity initialization and ReLU1 activations significantly improve robustness to quantization.

\begin{figure*}[t!]
    \begin{minipage}{\textwidth}%
        \begin{minipage}{0.35\textwidth}%
            \begin{subfigure}{\linewidth}%
                \centering
                \includegraphics[height=5cm]{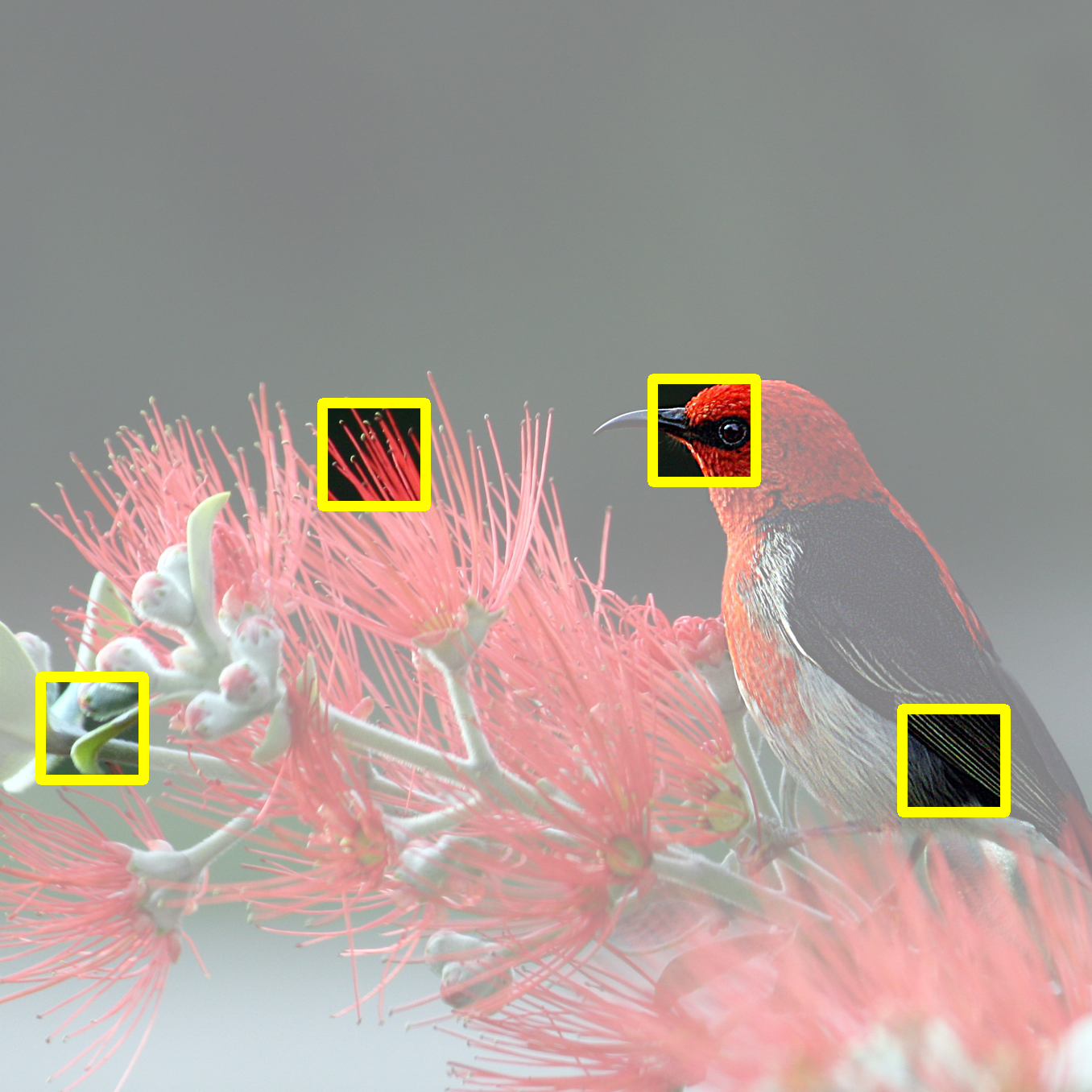}
                \vspace{-1mm}
                \caption*{}
                \label{fig:merge-scenario}
            \end{subfigure}
        \end{minipage}
        \begin{minipage}{0.8\textwidth}%
          \begin{subfigure}{0.2\linewidth}
            \centering
            \includegraphics[height=2.6cm]{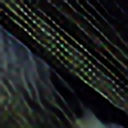}
            \vspace{-1mm}
            \caption*{FP16}
          \end{subfigure}
          \begin{subfigure}{0.2\linewidth}
            \centering
            \includegraphics[height=2.6cm]{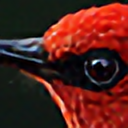}
            \vspace{-1mm}
            \caption*{FP16}
          \end{subfigure}
          \begin{subfigure}{0.2\linewidth}
            \centering
            \includegraphics[height=2.6cm]{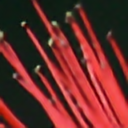}
            \vspace{-1mm}
            \caption*{FP16}
          \end{subfigure}
          \begin{subfigure}{0.2\linewidth}
            \centering
            \includegraphics[height=2.6cm]{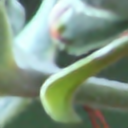}
            \vspace{-1mm}
            \caption*{FP16}
          \end{subfigure}
          
          \vspace{1mm}
          
          \begin{subfigure}{0.2\linewidth}
            \centering
            \includegraphics[height=2.6cm]{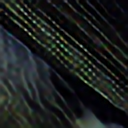}
            \vspace{-1mm}
            \caption*{INT8}
          \end{subfigure}
          \begin{subfigure}{0.2\linewidth}
            \centering
            \includegraphics[height=2.6cm]{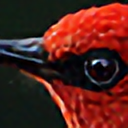}
            \vspace{-1mm}
            \caption*{INT8}
          \end{subfigure}
          \begin{subfigure}{0.2\linewidth}
            \centering
            \includegraphics[height=2.6cm]{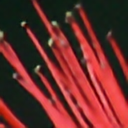}
            \vspace{-1mm}
            \caption*{INT8}
          \end{subfigure}
          \begin{subfigure}{0.2\linewidth}
            \centering
            \includegraphics[height=2.6cm]{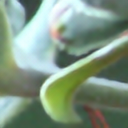}
            \vspace{-1mm}
            \caption*{INT8}
          \end{subfigure}
        \end{minipage}
    \end{minipage}
    \caption{Visual comparison of 4$\times$ super-resolved images from DIV2K produced by QuickSRNet-Medium before and after quantization.}
    \label{fig:fp16-vs-int8}
\end{figure*}

\subsection{Implementation details}

\paragraph{Baselines} We compare QuickSRNet against the following architectures: FSRCNN \cite{fsrcnn}, ESPCN \cite{espcn}, XLSR \cite{xlsr}, SESR \cite{sesr}, ABPN \cite{abpn}, ERFDN \cite{rfdn} and EDSR \cite{edsr}. Note that, rather than reporting PSNR and SSIM scores from the original papers, we re-implemented, trained and quantized all existing baselines from scratch. As a result, all models shared most hyper-parameters (batch-size, losses, optimizer, etc.), including the data loading/augmentation pipeline. We did however tweak the learning rate for each architecture independently. In some cases, our re-implementation deviates slightly from the original architecture when it includes operations that are not supported on the device used for profiling. For example, we replaced the parametric ReLUs \cite{prelu} used in SESR and FSRCNN to regular ReLUs. Despite these minor modifications, we were usually able to reproduce PSNR and SSIM scores reported in the original papers.

\paragraph{Training details} For most experiments, we train the models on the $800$ training images from the DIV2K dataset \cite{div2k} and evaluate them on standard SR testsets: Set5 \cite{set5}, Set14 \cite{set14}, BSD100 \cite{bsd100}, and Urban100 \cite{urban100}. We preprocess input and target images by scaling RGB values between $0$ and $1$. For data augmentation, we use random cropping, flipping and rotation. The models are trained for 1 million iterations with a batch size of $32$. We use an L1 loss and the Adam optimizer \cite{adam} with hyper-parameters $\epsilon=10^{-8}$ and $\beta=(0.9, 0.999)$. For the learning rate, we found that using an initial value of $5\times10^{-4}$ and decaying it by a factor of $0.5$ every $200$K iterations is a strategy that works well for most architectures. 
  
\paragraph{8-bit quantization} We use the AIMET library \cite{aimet} to perform model quantization \cite{nagel2021white} and compute post-quantization accuracy metrics\footnote{Additionally, we confirmed accuracy numbers on target for a subset of the models and typically found that the simulated numbers produced by AIMET to be within a $0.02$ range from the actual numbers obtained on target.}. Both weights and activations are quantized to 8-bit integers (W8A8 setup). We experimented with both Post-Training Quantization (PTQ) techniques and Quantization Aware Training (QAT). When we use QAT, we re-initialize the optimizer with a very small learning rate (usually $4\times10^{-6}$). 
  
\paragraph{On-device profiling} We profile the models on the Hexagon Processor of a device with Snapdragon 8 Gen 1 and report the average latency obtained on $100$ inputs of spatial resolution $512\times512$. Before profiling, the model is converted from PyTorch \cite{paszke2019pytorch} to ONNX. Please see the appendix for more details about the model conversion steps.

\begin{table*}[!th]
  \centering
  \setlength{\tabcolsep}{1em}
  \small
  \begin{tabular}{| l | c  c | c  c | c  c | c |}
    \hline
    \multirow{2}{*}{\textbf{QuickSRNet specs}} & \multicolumn{2}{c|}{\textbf{$\bm{2\times}$}} & \multicolumn{2}{c|}{\textbf{$\bm{3\times}$}} & \multicolumn{2}{c |}{\textbf{$\bm{4\times}$}} & \textbf{Latency} \\
    & \textbf{FP16} & \textbf{INT8} & \textbf{FP16} & \textbf{INT8} & \textbf{FP16} & \textbf{INT8} & \textbf{Measurements (ms)} \\
    \hline
    \textit{f32 - m1}           & 31.43 & 31.38 & 28.41 & 28.38 & 26.94 & 26.91 & 0.99 (\textcolor{medgreen}{$-$22\%}) \\
    \textit{f32 - m2} (small)   & 31.61 & 31.58 & 28.57 & 28.55 & 27.07 & 27.06 & 1.14 (\textcolor{medgreen}{$-$35\%}) \\
    \textit{f32 - m3}           & 31.72 & 31.63 & 28.67 & 28.63 & 27.16 & 27.12 & 1.21 (\textcolor{medgreen}{$-$34\%}) \\
    \textit{f32 - m5} (medium)  & 31.82 & 31.77 & 28.75 & 28.72 & 27.24 & 27.21 & 1.42 (\textcolor{medgreen}{$-$35\%}) \\
    \textit{f32 - m7}           & 31.88 & 31.81 & 28.81 & 28.76 & 27.30 & 27.27 & 1.74 (\textcolor{medgreen}{$-$30\%}) \\
    \textit{f32 - m11}          & 31.95 & 31.80 & 28.86 & 28.80 & 27.35 & 27.29 & 2.38 (\textcolor{medgreen}{$-$22\%}) \\
    \textit{f64 - m11} (large)  & 32.08 & 31.97 & 28.98 & 28.93 & 27.47 & 27.43 & 7.63 (\textcolor{medgreen}{\;\,$-$7\%}) \\
    \hline
  \end{tabular}
  \caption{PSNRs (dB) and latencies (ms) of various QuickSRNet configurations ($f$ : number of feature channels, $m$ : number of convolutional blocks in the network). We report PSNR numbers obtained before and after quantization. We also report latency measurements on a $512\times512$ input, obtained on a device with Snapdragon 8 Gen 1, and gains introduced by not using an input-to-output residual connection.}
  \label{table:qsrnet-psnr-numbers}
\end{table*}

\begin{table*}[!th]
  \centering
  \setlength{\tabcolsep}{1em}
  \small
  \begin{tabular}{| l | c  c | c  c | c  c | c |}
    \hline
    \multirow{2}{*}{\textbf{Existing Models}} & \multicolumn{2}{c|}{\textbf{$\bm{2\times}$}} & \multicolumn{2}{c|}{\textbf{$\bm{3\times}$}} & \multicolumn{2}{c |}{\textbf{$\bm{4\times}$}} & \textbf{Latency} \\
    & \textbf{FP16} & \textbf{INT8} & \textbf{FP16} & \textbf{INT8} & \textbf{FP16} & \textbf{INT8} & \textbf{Measurements (ms)} \\
    \hline
    XLSR                       & 31.62 & 31.32 & 28.59 & 28.31 & 27.09 & 26.82 &  1.59 \\
    ESPCN                      & 31.37 & 30.19 & 28.35 & 27.86 & 26.87 & 26.44 &  1.83 \\
    SESR-M3                    & 31.57 & 31.40 & 28.52 & 28.47 & 27.02 & 26.97 &  2.02 \\
    ABPN                       & 31.80 & 31.74 & 28.73 & 28.70 & 27.22 & 27.20 &  2.09 \\
    SESR-M5                    & 31.68 & 31.53 & 28.63 & 28.56 & 27.11 & 27.05 &  2.23 \\
    SESR-M7                    & 31.76 & 31.67 & 28.68 & 28.61 & 27.16 & 27.07 &  2.33 \\
    SESR-M11                   & 31.84 & 31.74 & 28.77 & 28.60 & 27.25 & 27.18 &  3.06 \\
    FSRCNN                     & 31.50 & 31.22 & 28.49 & 28.32 & 26.99 & 26.89 &  3.45 \\
    SESR-XL                    & 32.02 & 31.94 & 28.91 & 28.85 & 27.39 & 27.35 &  3.75 \\
    ERFDN                      & 32.20 & 32.06 & 29.08 & 28.98 & 27.57 & 27.48 & 19.50  \\
    EDSR                       & 32.21 & 32.08 & 29.04 & 28.89 & 27.61 & 27.53 & 37.95 \\
    \hline
  \end{tabular}
  \caption{PSNRs (dB) and latencies (ms) of existing SISR solutions on BSD100. Please note that we re-implemented, trained, and quantized all architectures from scratch. Latency numbers were measured on a device with Snapdragon 8 Gen 1, using a $512\times512$ input.}
  \label{table:sisr-baselines-psnr-numbers}
\end{table*}

\begin{figure}[t!]
  \centering
  \begin{subfigure}{0.48\linewidth}
    \centering
    \includegraphics[width=\linewidth]{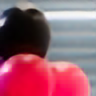}
    \caption*{ABPN}
  \end{subfigure}
  \hspace{1mm}
  \begin{subfigure}{0.48\linewidth}
    \centering
    \includegraphics[width=\linewidth]{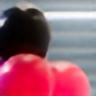}
    \caption*{QSRNet-Medium}
  \end{subfigure}
  \caption{Visual artifacts by ABPN vs QuickSRNet-Medium on $4\times$-upscaled images from Urban100. More examples can be found in the supplementary material.}
  \label{fig:abpn-vs-qsrnet-artifacts}
\end{figure}

\begin{figure*}[!t]
\centering
\includegraphics[width=0.9\linewidth]{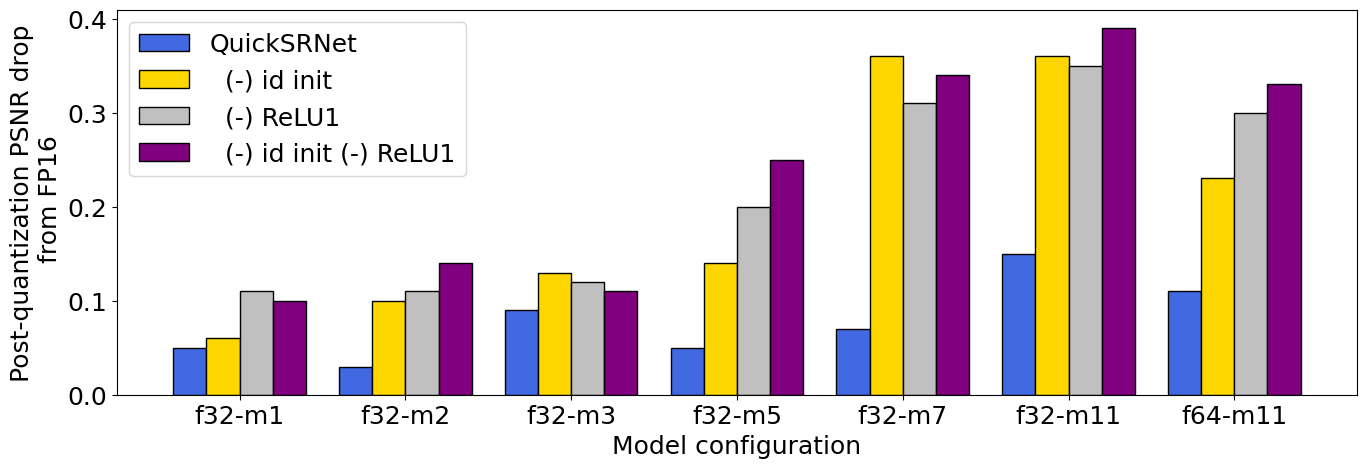}
\caption{Ablation study comparing the post-quantization PSNR drop from FP16 when removing identity initialization and/or ReLU1 activations from the architecture design.}
\label{fig:qsrnet-psnr-drop-from-fp16}
\end{figure*}

\begin{table*}[ht!]
\centering
\setlength{\tabcolsep}{0.5em}
\small
\begin{tabular}{|c|c|c|c|c|c|}
\hline
\multirow{2}{*}{\textbf{Specification}}     & \textbf{Post-training} & \textbf{No optimizations}  & \textbf{QAT}  & \textbf{Per-channel QAT}  & \textbf{Per-channel Adaround}                                                                       \\
                                            & \textbf{FP16}          & \textbf{INT8}                                       & \textbf{INT8}                                        & \textbf{INT8}                                                 & \textbf{INT8}                                                 \\ \hline
\textit{QuickSRNet-Small}  & 31.61                  & 30.81 (\textcolor{red}{$-$0.80}) & 31.34 (\textcolor{red}{$-$0.27})  & \textbf{31.57 (\textcolor{gray}{$-$0.04})}          & 31.56 (\textcolor{gray}{$-$0.05}) \\
\textit{QuickSRNet-Medium} & 31.82                  & 30.74 (\textcolor{red}{$-$1.08}) & 31.61 (\textcolor{red}{$-$0.21})  & 31.75 (\textcolor{gray}{$-$0.07}) & \textbf{31.77 (\textcolor{gray}{$-$0.05})}          \\
\textit{QuickSRNet-Large}  & 32.07                  & 31.37 (\textcolor{red}{$-$0.70}) & 31.90 (\textcolor{gray}{$-$0.10}) & 31.97 (\textcolor{gray}{$-$0.10}) & \textbf{31.99 (\textcolor{gray}{$-$0.08})}          \\ \hline
\end{tabular}
\caption{Impact of various quantization techniques on accuracy. Activations are always quantized to 8-bit integers using per-tensor quantization. For weights, we tried both per-tensor and per-channel quantization and found the latter to work significantly better.}
\label{table:quant-experiments}
\end{table*}

\begin{figure*}[th!]
  \centering
  \begin{subfigure}{0.2\linewidth}
    \centering
    \includegraphics[width=\linewidth]{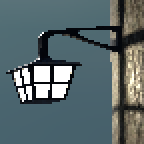}
    \caption{LR}
  \end{subfigure}
  \hspace{1mm}
  \begin{subfigure}{0.2\linewidth}
    \centering
    \includegraphics[width=\linewidth]{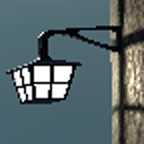}
    \caption{Bicubic}
  \end{subfigure}
  \hspace{1mm}
  \begin{subfigure}{0.2\linewidth}
    \centering
    \includegraphics[width=\linewidth]{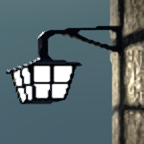}
    \caption{FSR1.0}
  \end{subfigure}
  \hspace{1mm}
  \begin{subfigure}{0.2\linewidth}
    \centering
    \includegraphics[width=\linewidth]{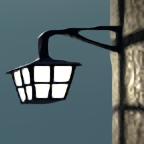}
    \caption{QuickSRNet-Small}
  \end{subfigure}
  \hfill
  \caption{SISR ($2\times$) for Gaming: (a) Low-resolution, (b) Bicubic interpolation, (c) FSR1.0 \cite{fsr}, and (d) QuickSRNet-Small (ours)}
  \label{fig:fsr-vs-qsr}
\end{figure*}

\begin{table}[!t]
  \centering

  \setlength{\tabcolsep}{1em}
  \small
  \begin{tabular}{| c | c | c |}
    \hline
    \textbf{Method} & \textbf{PSNR} & \textbf{SSIM} \\
    \hline
    \textit{Bicubic} & 28.88 & 0.8683 \\
    \textit{FSR1.0} & 29.01 & 0.8707 \\
    \textit{QuickSRNet-Small} & 29.71 & 0.8806 \\
    \hline
  \end{tabular}
  \caption{PSNR/SSIM scores for different $2\times$ single-image super-resolution solutions for gaming.}
  \label{table:qsrnet-vs-fsr-gaming-numbers}
\end{table}

\section{Experimental results}
\label{sec:experiments}

In this section, we compare QuickSRNet against existing SR architectures in terms of accuracy-to-latency trade-offs and demonstrate the effectiveness of our training tricks to improve robustness to quantization through ablation studies.  

\paragraph{Scaling laws of QuickSRNet} We experimented with several architecture specifications, varying the number of conv modules $m$ and the number of feature channels $f$. PSNR and SSIM scores on the BSD100 dataset obtained with each specification and a scaling factor of $2$ can be found in \cref{table:qsrnet-psnr-numbers}. As expected, larger/wider networks obtain higher PSNR/SSIM scores. The measured latency for each specification is reported on the last column and we indicate for each architecture the latency improvement introduced by removing the input-to-output connection. In the rest of the paper, we only use a subset of these model configurations: QuickSRNet-small (a.k.a. \textit{QuickSRNet-f32-m2}), QuickSRNet-medium (a.k.a. \textit{QuickSRNet-f32-m5}), and QuickSRNet-large (a.k.a. \textit{QuickSRNet-f64-m11}). As can be seen in \cref{fig:quicksrnet-perf} and \cref{table:qsrnet-psnr-numbers,table:sisr-baselines-psnr-numbers}, these $3$ variants obtain similar accuracy scores in contrast to SESR-M7, ABPN and EDSR respectively while being significantly faster. Please refer to \cref{fig:sr-comp-one-col} for a comparison of QuickSRNet-\{small, medium, large\} versus SESR-M7, ABPN and EDSR in terms of image quality and latency improvement. More  quantitative results and side-by-side comparisons for $2\times$, $3\times$ and $4\times$ upscaling are available in the supplementary materials.

\paragraph{W8A8 quantization} For all our experiments, we quantize both, model weights and activations, to 8-bit integers. Without any optimizations, we observe a significant drop after quantization (see \cref{table:quant-experiments}). While finetuning the per-tensor quantized weights via QAT can recover some of this drop, we found per-channel weight quantization to be important.

Furthermore, we experimented with several post-training quantization methods, including: cross-layer equalization (CLE) \cite{clebc}, bias correction (BC) \cite{clebc}, and adaptive rounding (AdaRound) \cite{adaround}, and found AdaRound to obtain comparable performance to per-channel QAT, outperforming the other PTQ approaches. CLE did not work well in our experiments, most likely because it skews the activation values outside the ReLU1 range. In an attempt to further improve post-quantization accuracy, we tried finetuning the per-channel Adarounded weights using QAT but this did not improve post-quantization accuracy. 


\paragraph{Robustness to quantization} Overall, QuickSRNet quantizes well to W8A8. As can be seen in \cref{fig:fp16-vs-int8}, images produced by the quantized model are indistinguishable from their full-precision counterparts. In \cref{fig:qsrnet-psnr-drop-from-fp16}, we visualize the drop in PSNR post quantization and show the benefits of combining identity initialization and ReLU1 activations. Regardless of the model size, removing one or both of these ingredients from the model design results in a significantly worse accuracy after quantization. 

\paragraph{Less prone to block artifacts} Our experiments show that architectures with a nearest-neighbour upsampling skip connection tend to produce outputs with block-like artifacts of size $S \times S$. Interestingly, our residual-free architecture seems less prone to this issue and produces more perceptually pleasing results. A visual comparison of such artifacts can be seen in \cref{fig:abpn-vs-qsrnet-artifacts}.

\begin{table}[!t]
  \centering
  \setlength{\tabcolsep}{0.5em}
  \small
  \begin{tabular}{| c | c | c | c | c | c |}
    \hline
    \textbf{Target resolution} & \textit{540p} & \textit{720p} & \textit{1080p} & \textit{1440p} & \textit{2160p} \\
    \hline
    \textbf{Latency (ms)} & 0.69 & 0.95 & 2.24 & 4.25 & 8.15 \\
    \hline
  \end{tabular}
  \caption{QuickSRNet-Small latency (ms) running at different target resolutions on a device with Snapdragon 8 Gen 1.}
  \label{table:qsrnet-different-resolution-gaming-latency}
\end{table}

\begin{figure*}[th!]
  \centering
  \begin{subfigure}{0.48\linewidth}
    \centering
    \includegraphics[width=\linewidth]{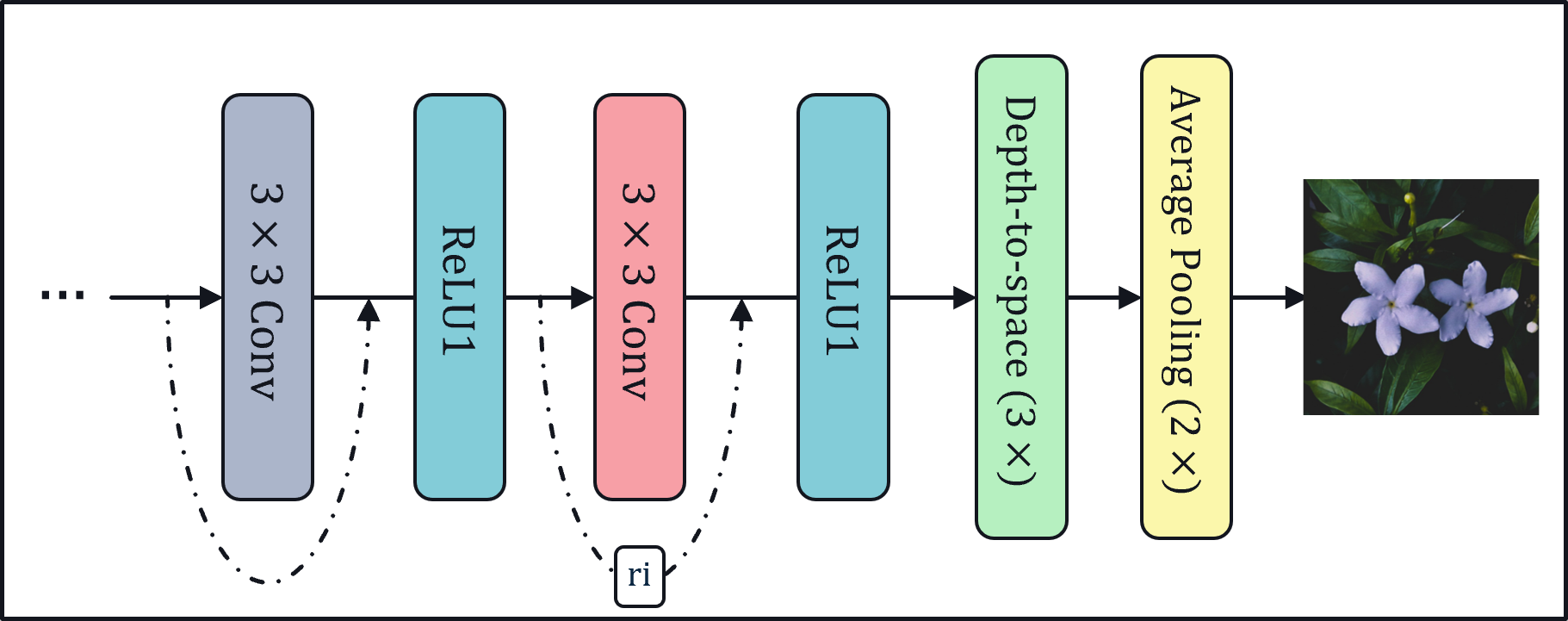}
    \caption{Na\"{i}ve baseline}
  \end{subfigure}
  \hspace{1mm}
  \begin{subfigure}{0.48\linewidth}
    \centering
    \includegraphics[width=\linewidth]{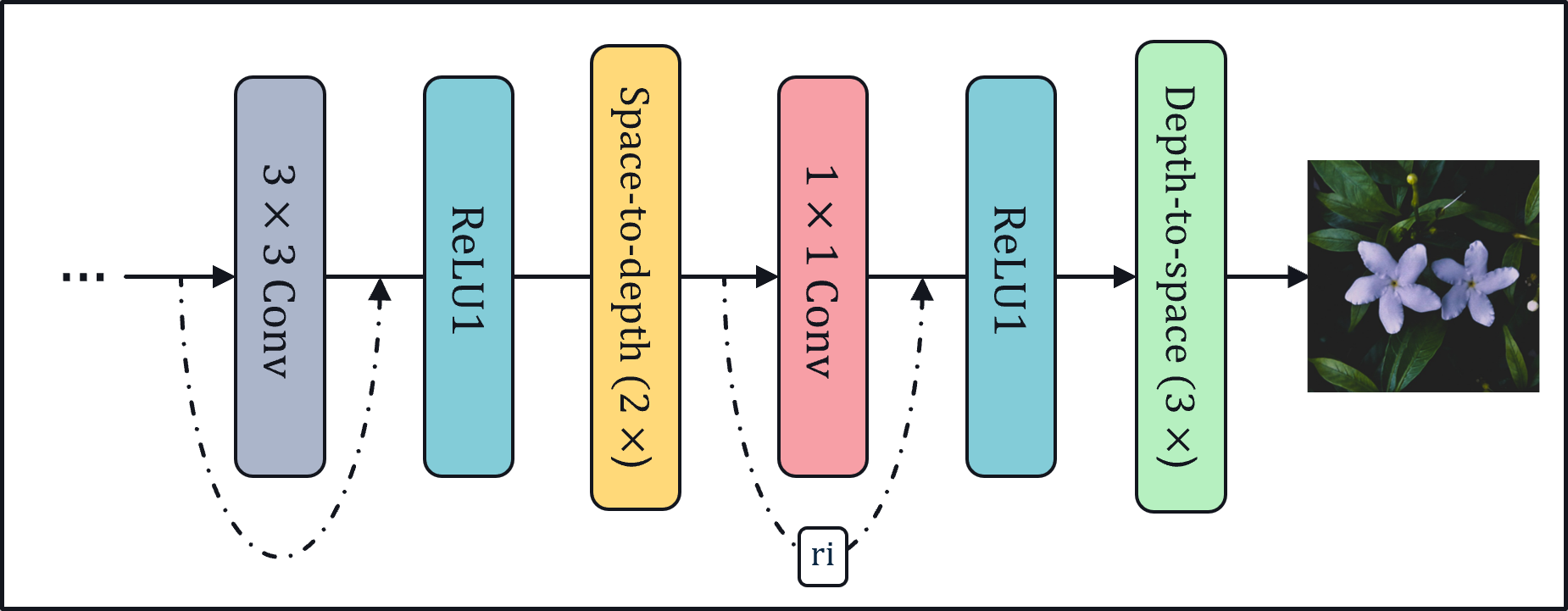}
    \caption{Proposed approach}
  \end{subfigure}
  \caption{Two different architecture modifications to implement $1.5\times$ upscaling: (a) Na\"{i}ve approach, where we repurpose a $3\times$ architecture by adding an average pooling layer on top, (b) Our approach, where we halve the resolution inside the network and map to target resolution using a $3\times$ subpixel conv.}
  \label{fig:qsrnet-1.5x-models}
\end{figure*}

\section{DL-based SISR for mobile gaming}

A real-world application of efficient super-resolution is video gaming. While DL-based super-resolution (or supersampling) has already been commericalized on high-end gaming desktops \cite{dlss, xess}, these solutions are not supported on mobile platforms yet. One specificity of gaming content is that synthetically rendered images are significantly more aliased than natural images. Nevertheless, we find \textit{QuickSRNet-Small} to work well on this domain, with no changes needed apart from re-training it on gaming data.
\Cref{fig:fsr-vs-qsr} shows some results obtained by \textit{QuickSRNet-Small} when applied to gaming content. We compare our results against non-ML based single-frame upscaling approaches, including an FSR1.0 baseline \cite{fsr} which was specifically designed for this use case. Overall, we find that \textit{QuickSRNet-Small} produces better-looking images compared to the other baselines. The visual benefits also translate into PSNR and SSIM gains, as can be seen in \cref{table:qsrnet-vs-fsr-gaming-numbers}. In terms of latency, \cref{table:qsrnet-different-resolution-gaming-latency} shows \textit{QuickSRNet-Small} latency measurements at various target resolutions, from $540$p to $4$k. In the future, we would like to extend our architecture to the multi-frame case which has become the de facto standard for video gaming (\eg FSR 2.0, \cite{fsr2_2022}, DLSS 2.0 \cite{liu2020dlss}, XeSS \cite{chowdhury2022intel}).  

\begin{table}[t!]
    \centering
    \setlength{\tabcolsep}{1em}
    \small
    \begin{tabular}{| c | c | c | c |}
        \hline
        \textbf{QuickSRNet} & \multirow{2}{*}{\textbf{Bicubic}} & \textbf{Na\"{i}ve} & \textbf{Proposed} \\
        \textbf{Specification} &  & \textbf{Baseline} & \textbf{Approach} \\
        \hline
        \textit{Small} & \multirow{3}{*}{32.47} & 34.71 & 34.89 \\
        \textit{Medium} &  & 34.87 & 35.13 \\
        \textit{Large} &  & 35.18 & 35.47 \\
        \hline
    \end{tabular}
    \caption{PSNRs (dB) evaluated after quantization on BSD100 dataset via $1.5\times$ upscaling.}
    \label{table:qsrnet-1.5x-psnr-numbers}
\end{table}



\subsection{QuickSRNet \texorpdfstring{$\bm{1.5\times}$}{1.5x}}

Standard super-resolution datasets are usually limited to $2\times$, $3\times$ or $4\times$ upscaling and non-integer scaling factors are rarely explored. On the other hand, $1.5 \times$ upscaling is often proposed in VR and gaming applications\footnote{Both DLSS and FSR support $1.5\times$ via their ``Quality" mode.}. In this section, we describe an approach to perform $1.5\times$ upscaling, a setting that is not trivially supported by most efficient SR architectures as non-integer scaling factors are not compatible with the final sub-pixel convolution.

\paragraph{$\bm{3\times}$ upscaling followed by $\bm{2\times}$ downscaling baseline} A na\"{i}ve approach to $1.5\times$ upscaling consists in downscaling by a factor $2$ the output of a $3\times$ SR model. This can be achieved by adding a $2\times2$ average pooling layer at the end of the architecture.

\paragraph{Proposed $\bm{1.5\times}$ upscaling approach} Instead, we propose to halve the resolution inside the network using a \textit{space-to-depth} operation with a block-size of $2$ which we then map to target resolution using a $3\times$ subpixel convolution. To compensate for the $4\times$ increase of channels due to the \textit{space-to-depth} operation, we implement the subpixel convolution using a $1\times1$ kernel.

\paragraph{} \Cref{fig:qsrnet-1.5x-models} shows the two considered $1.5\times$ architecture heads. As can be seen in \cref{table:qsrnet-1.5x-psnr-numbers}, the proposed approach significantly outperforms the na\"{i}ve  $3\times$ upscaling followed by $2\times$ downscaling baseline.

\section{Conclusion}

\paragraph{} In this study, we propose QuickSRNet, an efficient super-resolution architecture for mobile platforms. We have thoroughly analyzed the performance of our models and existing ones, systematically checking accuracy after quantization and profiling latency on a mobile device. Our experiments have shown that QuickSRNet is well suited for real-time applications on mobile devices due to its high speed and good accuracy. We have also demonstrated the effectiveness of our solution on a real world use case (mobile gaming) and believe that our training tricks to improve robustness to quantization are applicable to other works. We have released the implementation and pretrained weights (including quantized weights) of QuickSRNet models as part of the \hyperlink{https://github.com/quic/aimet-model-zoo}{AIMET model zoo}\footnote{For QuickSRNet-large, the released version of the model includes the input-to-output residual connection as this leads to slightly higher accuracy and the latency improvement (-7\%) is minimal for larger architectures.}. We believe that QuickSRNet provides a practical solution for applications that require real-time super-resolution capabilities.

{\small
\bibliographystyle{ieee_fullname}
\bibliography{egbib}
}


\newpage
\onecolumn

\section*{Supplementary Material}

\begin{figure*}[!h]
    \begin{minipage}{\textwidth}%
        
        \begin{minipage}{0.35\textwidth}%
            \begin{subfigure}{\linewidth}%
                \centering
                \includegraphics[height=5cm]{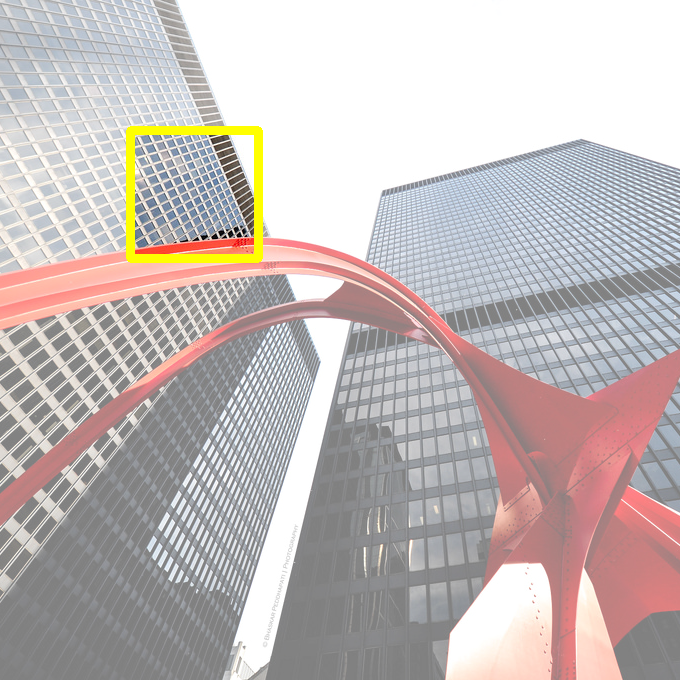}
                \vspace{-1mm}
                \caption*{}
                \label{fig:merge-scenario}
            \end{subfigure}
        \end{minipage}
        \begin{minipage}{0.8\textwidth}%
          \begin{subfigure}{0.2\linewidth}
            \centering
            \includegraphics[height=2.6cm]{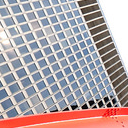}
            \vspace{-1mm}
            \caption*{HR}
          \end{subfigure}
          \begin{subfigure}{0.2\linewidth}
            \centering
            \includegraphics[height=2.6cm]{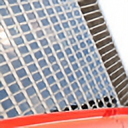}
            \vspace{-1mm}
            \caption*{SESR-M7}
          \end{subfigure}
          \begin{subfigure}{0.2\linewidth}
            \centering
            \includegraphics[height=2.6cm]{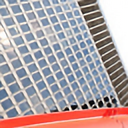}
            \vspace{-1mm}
            \caption*{ABPN}
          \end{subfigure}
          \begin{subfigure}{0.2\linewidth}
            \centering
            \includegraphics[height=2.6cm]{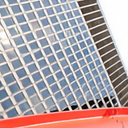}
            \vspace{-1mm}
            \caption*{EDSR}
          \end{subfigure}
          
          \vspace{1mm}
          
          \begin{subfigure}{0.2\linewidth}
            \centering
            \includegraphics[height=2.6cm]{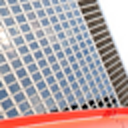}
            \vspace{-1mm}
            \caption*{Bicubic}
          \end{subfigure}
          \begin{subfigure}{0.2\linewidth}
            \centering
            \includegraphics[height=2.6cm]{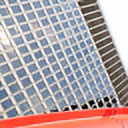}
            \vspace{-1mm}
            \caption*{Ours-Small}
          \end{subfigure}
          \begin{subfigure}{0.2\linewidth}
            \centering
            \includegraphics[height=2.6cm]{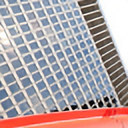}
            \vspace{-1mm}
            \caption*{Ours-Medium}
          \end{subfigure}
          \begin{subfigure}{0.2\linewidth}
            \centering
            \includegraphics[height=2.6cm]{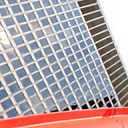}
            \vspace{-1mm}
            \caption*{Ours-Large}
          \end{subfigure}
        \end{minipage}
    \caption{Visual comparison of $2\times$ super-resolution results by QuickSRNet and existing solutions on Urban100 images.}
    \label{fig:2x-results}
    \end{minipage}%
    
    \vspace{0.2cm}
    
    \begin{minipage}{\textwidth}%
        
        \begin{minipage}{0.35\textwidth}%
            \begin{subfigure}{\linewidth}%
                \centering
                \includegraphics[height=5cm]{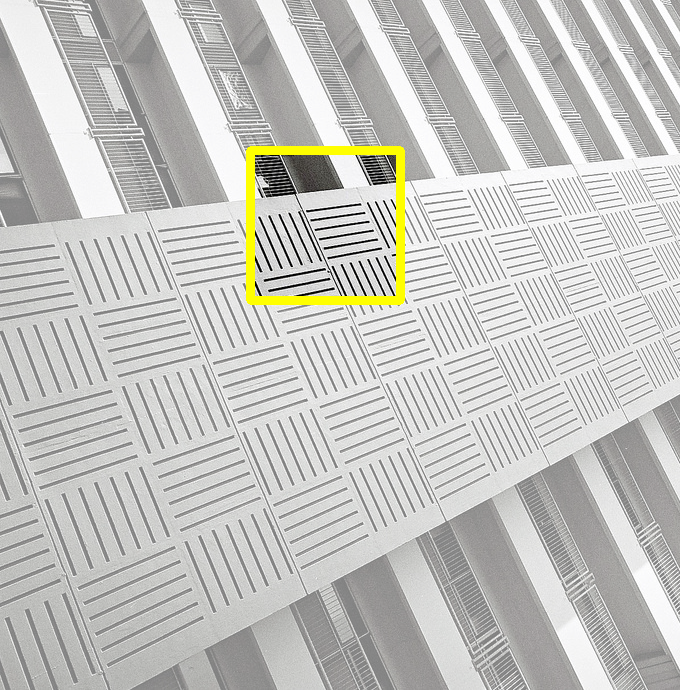}
                \vspace{-1mm}
                \caption*{}
                \label{fig:merge-scenario}
            \end{subfigure}
        \end{minipage}
        \begin{minipage}{0.8\textwidth}%
          \begin{subfigure}{0.2\linewidth}
            \centering
            \includegraphics[height=2.6cm]{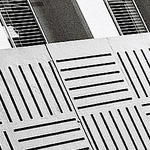}
            \vspace{-1mm}
            \caption*{HR}
          \end{subfigure}
          \begin{subfigure}{0.2\linewidth}
            \centering
            \includegraphics[height=2.6cm]{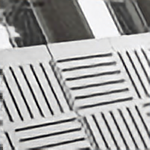}
            \vspace{-1mm}
            \caption*{SESR-M7}
          \end{subfigure}
          \begin{subfigure}{0.2\linewidth}
            \centering
            \includegraphics[height=2.6cm]{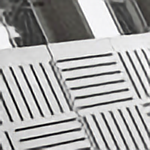}
            \vspace{-1mm}
            \caption*{ABPN}
          \end{subfigure}
          \begin{subfigure}{0.2\linewidth}
            \centering
            \includegraphics[height=2.6cm]{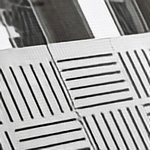}
            \vspace{-1mm}
            \caption*{EDSR}
          \end{subfigure}
          
          \vspace{1mm}
          
          \begin{subfigure}{0.2\linewidth}
            \centering
            \includegraphics[height=2.6cm]{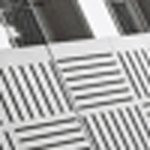}
            \vspace{-1mm}
            \caption*{Bicubic}
          \end{subfigure}
          \begin{subfigure}{0.2\linewidth}
            \centering
            \includegraphics[height=2.6cm]{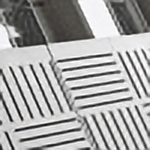}
            \vspace{-1mm}
            \caption*{Ours-Small}
          \end{subfigure}
          \begin{subfigure}{0.2\linewidth}
            \centering
            \includegraphics[height=2.6cm]{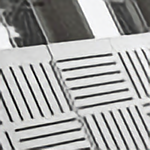}
            \vspace{-1mm}
            \caption*{Ours-Medium}
          \end{subfigure}
          \begin{subfigure}{0.2\linewidth}
            \centering
            \includegraphics[height=2.6cm]{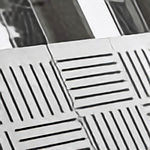}
            \vspace{-1mm}
            \caption*{Ours-Large}
          \end{subfigure}
        \end{minipage}
        
    \caption{Visual comparison of $3\times$ super-resolution results by QuickSRNet and existing solutions on Urban100 images.}
    \label{fig:3x-results}
    \end{minipage}%
    
    \vspace{0.2cm}

    \begin{minipage}{\textwidth}%
        
        \begin{minipage}{0.35\textwidth}%
            \begin{subfigure}{\linewidth}%
                \centering
                \includegraphics[height=5cm]{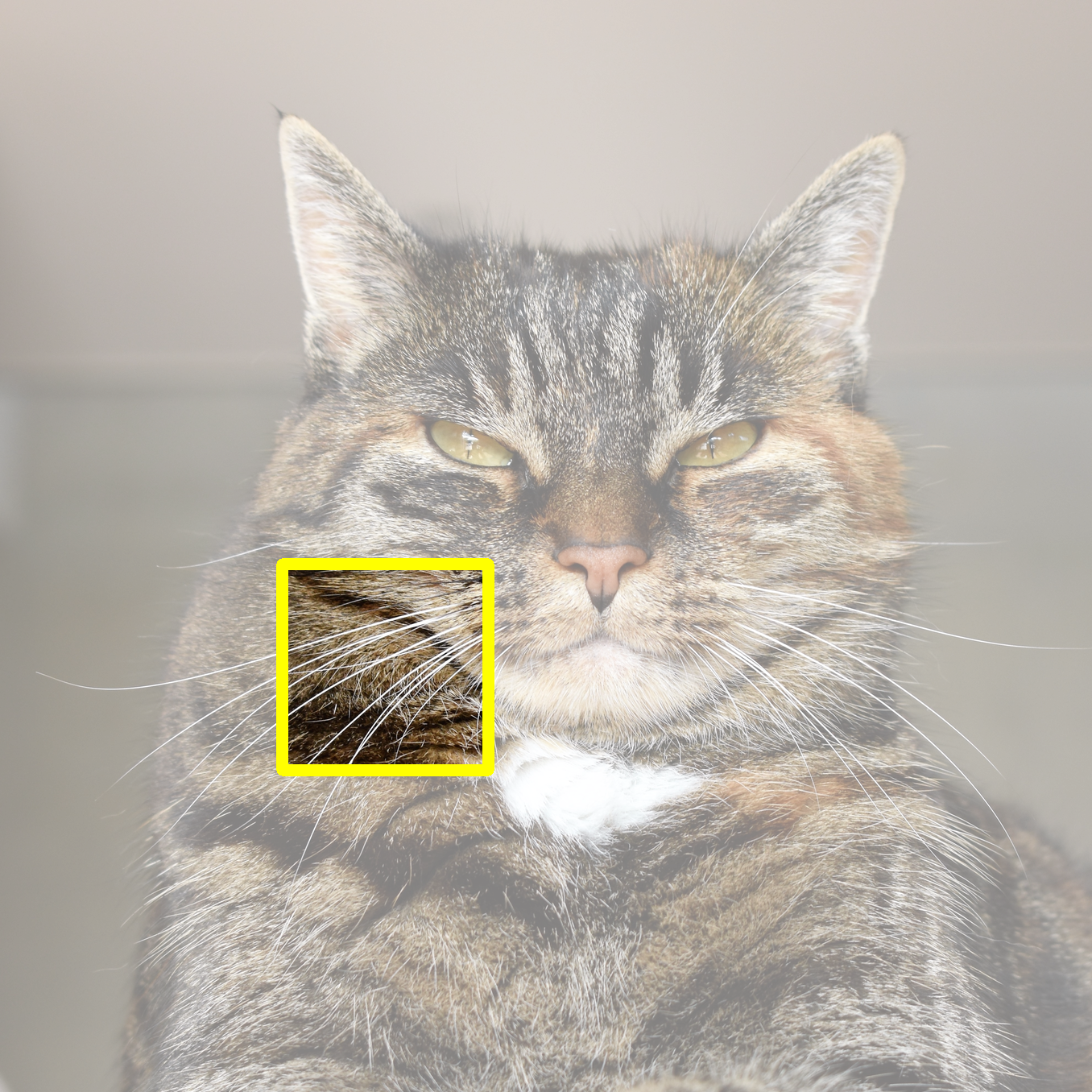}
                \vspace{-1mm}
                \caption*{}
                \label{fig:merge-scenario}
            \end{subfigure}
        \end{minipage}
        \begin{minipage}{0.8\textwidth}%
          \begin{subfigure}{0.2\linewidth}
            \centering
            \includegraphics[height=2.6cm]{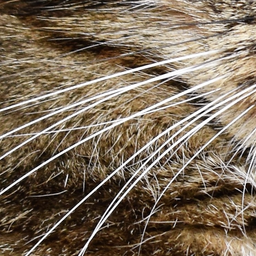}
            \vspace{-1mm}
            \caption*{HR}
          \end{subfigure}
          \begin{subfigure}{0.2\linewidth}
            \centering
            \includegraphics[height=2.6cm]{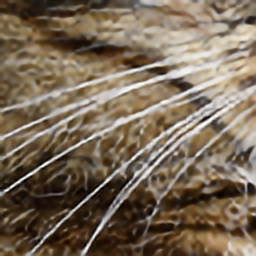}
            \vspace{-1mm}
            \caption*{SESR-M7}
          \end{subfigure}
          \begin{subfigure}{0.2\linewidth}
            \centering
            \includegraphics[height=2.6cm]{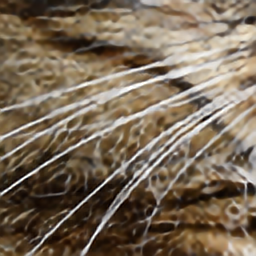}
            \vspace{-1mm}
            \caption*{ABPN}
          \end{subfigure}
          \begin{subfigure}{0.2\linewidth}
            \centering
            \includegraphics[height=2.6cm]{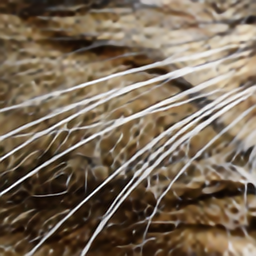}
            \vspace{-1mm}
            \caption*{EDSR}
          \end{subfigure}
          
          \vspace{1mm}
          
          \begin{subfigure}{0.2\linewidth}
            \centering
            \includegraphics[height=2.6cm]{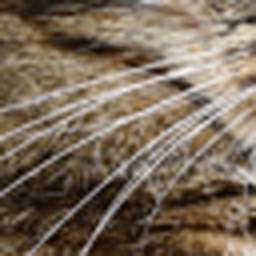}
            \vspace{-1mm}
            \caption*{Bicubic}
          \end{subfigure}
          \begin{subfigure}{0.2\linewidth}
            \centering
            \includegraphics[height=2.6cm]{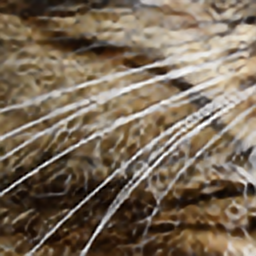}
            \vspace{-1mm}
            \caption*{Ours-Small}
          \end{subfigure}
          \begin{subfigure}{0.2\linewidth}
            \centering
            \includegraphics[height=2.6cm]{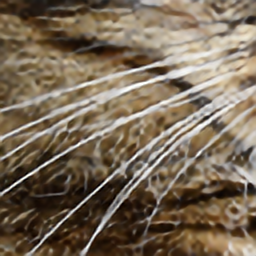}
            \vspace{-1mm}
            \caption*{Ours-Medium}
          \end{subfigure}
          \begin{subfigure}{0.2\linewidth}
            \centering
            \includegraphics[height=2.6cm]{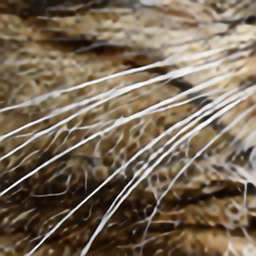}
            \vspace{-1mm}
            \caption*{Ours-Large}
          \end{subfigure}
        \end{minipage}
    \caption{Visual comparison of $4\times$ super-resolution results by QuickSRNet and existing solutions on DIV2K images.}
    \label{fig:4x-results}
    \end{minipage}%

\end{figure*}

\begin{figure*}[h!]
  \centering
  \begin{minipage}{0.48\linewidth}
      \begin{subfigure}{0.48\linewidth}
        \centering
        \includegraphics[width=\linewidth]{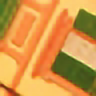}
        \caption*{ABPN}
      \end{subfigure}
      \hspace{0.25mm}
      \begin{subfigure}{0.48\linewidth}
        \centering
        \includegraphics[width=\linewidth]{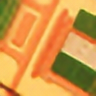}
        \caption*{QuickSRNet-Medium}
      \end{subfigure}
  \end{minipage}
  \hspace{4mm}
  \begin{minipage}{0.48\linewidth}
      \begin{subfigure}{0.48\linewidth}
        \centering
        \includegraphics[width=\linewidth]{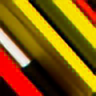}
        \caption*{ABPN}
      \end{subfigure}
      \hspace{0.25mm}
      \begin{subfigure}{0.48\linewidth}
        \centering
        \includegraphics[width=\linewidth]{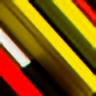}
        \caption*{QuickSRNet-Medium}
      \end{subfigure}
  \end{minipage}
  \caption{More examples of visual artifacts by ABPN vs QuickSRNet-Medium ($4\times$) on Urban 100 images.}
  \label{fig:supplementary-abpn-vs-qsrnet-artifacts}
\end{figure*}

\begin{figure*}[h!]
  \centering
  \begin{subfigure}{0.23\linewidth}
    \centering
    \includegraphics[width=\linewidth]{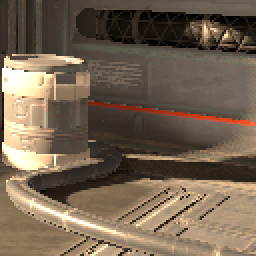}
  \end{subfigure}
  \hspace{1mm}
  \begin{subfigure}{0.23\linewidth}
    \centering
    \includegraphics[width=\linewidth]{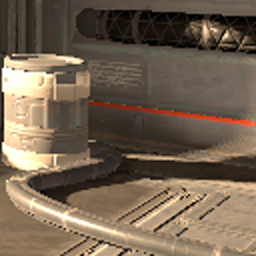}
  \end{subfigure}
  \hspace{1mm}
  \begin{subfigure}{0.23\linewidth}
    \centering
    \includegraphics[width=\linewidth]{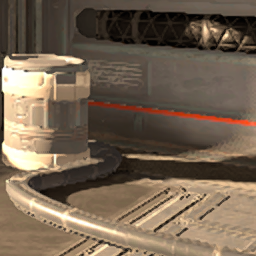}
  \end{subfigure}
  \hspace{1mm}
  \begin{subfigure}{0.23\linewidth}
    \centering
    \includegraphics[width=\linewidth]{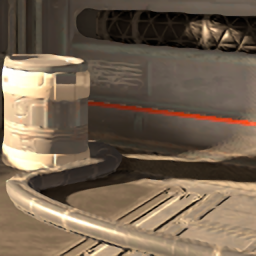}
  \end{subfigure}
  \hfill
  \begin{subfigure}{0.23\linewidth}
    \centering
    \includegraphics[width=\linewidth]{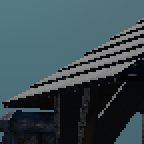}
  \end{subfigure}
  \hspace{1mm}
  \begin{subfigure}{0.23\linewidth}
    \centering
    \includegraphics[width=\linewidth]{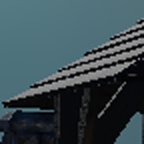}
  \end{subfigure}
  \hspace{1mm}
  \begin{subfigure}{0.23\linewidth}
    \centering
    \includegraphics[width=\linewidth]{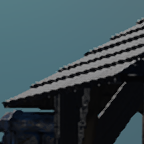}
  \end{subfigure}
  \hspace{1mm}
  \begin{subfigure}{0.23\linewidth}
    \centering
    \includegraphics[width=\linewidth]{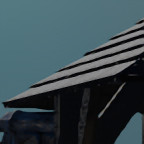}
  \end{subfigure}
  \begin{subfigure}{0.23\linewidth}
    \centering
    \includegraphics[width=\linewidth]{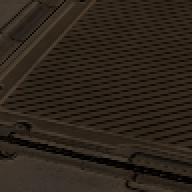}
    \caption{LR}
  \end{subfigure}
  \hspace{1mm}
  \begin{subfigure}{0.23\linewidth}
    \centering
    \includegraphics[width=\linewidth]{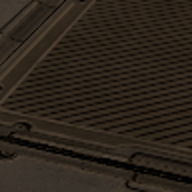}
    \caption{Bicubic}
  \end{subfigure}
  \hspace{1mm}
  \begin{subfigure}{0.23\linewidth}
    \centering
    \includegraphics[width=\linewidth]{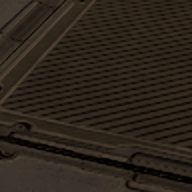}
    \caption{FSR1.0}
  \end{subfigure}
  \hspace{1mm}
  \begin{subfigure}{0.23\linewidth}
    \centering
    \includegraphics[width=\linewidth]{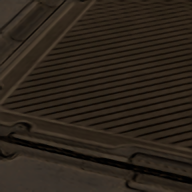}
    \caption{QuickSRNet-Small}
  \end{subfigure}

  \caption{SISR ($2\times$) for Gaming: (a) Low-resolution, (b) Bicubic interpolation, (c) FSR1.0, and (d) QuickSRNet-Small (ours).}
  \label{fig:supplementary-fsr-vs-qsr}
\end{figure*}

\newpage

\begin{table*}[t]
  \centering
  \setlength{\tabcolsep}{1em}
  \footnotesize
  \begin{tabular}{| c | l | c  c | c  c | c  c | c  c |}
    \hline
    \textbf{Scaling} & \multirow{2}{*}{\textbf{QuickSRNet Specification}} & \multicolumn{2}{c|}{\textbf{Set5}} & \multicolumn{2}{c|}{\textbf{Set14}} & \multicolumn{2}{c|}{\textbf{BSD100}} & \multicolumn{2}{c|}{\textbf{Urban100}} \\
    \textbf{Factor} &  & \textbf{FP16} & \textbf{INT8} & \textbf{FP16} & \textbf{INT8} & \textbf{FP16} & \textbf{INT8} & \textbf{FP16} & \textbf{INT8} \\
    \hline
    \multirow{8}{*}{2$\times$} & \textit{f32 - m1} & 36.83 & 36.67 & 32.35 & 32.28 & 31.43 & 31.38 & 29.66 & 29.61 \\
    & \textit{f32 - m2} (small)                    & 37.12 & 36.97 & 32.57 & 32.53 & 31.61 & 31.58 & 30.15 & 30.10 \\
    & \textit{f32 - m3}                            & 37.30 & 37.06 & 32.72 & 32.57 & 31.72 & 31.63 & 30.43 & 30.30 \\
    & \textit{f32 - m5} (medium)                   & 37.39 & 37.22 & 32.82 & 32.75 & 31.82 & 31.77 & 30.75 & 30.66 \\
    & \textit{f32 - m7}                            & 37.51 & 37.27 & 32.95 & 32.84 & 31.88 & 31.81 & 30.93 & 30.84 \\
    & \textit{f32 - m11}                           & 37.59 & 37.19 & 33.00 & 32.86 & 31.95 & 31.80 & 31.14 & 30.91 \\
    & \textit{f64 - m11} (large)                   & 37.87 & 37.61 & 33.29 & 33.18 & 32.12 & 32.04 & 31.74 & 31.64 \\
    \hline
    \multirow{8}{*}{3$\times$} & \textit{f32 - m1} & 32.75 & 32.69 & 29.08 & 29.05 & 28.41 & 28.38 & 26.19 & 26.16 \\
    & \textit{f32 - m2} (small)                    & 33.10 & 33.03 & 29.29 & 29.25 & 28.57 & 28.55 & 26.53 & 26.51 \\
    & \textit{f32 - m3}                            & 33.33 & 33.25 & 29.39 & 29.35 & 28.67 & 28.63 & 26.77 & 26.72 \\
    & \textit{f32 - m5} (medium)                   & 33.58 & 33.49 & 29.49 & 29.46 & 28.75 & 28.72 & 27.02 & 26.99 \\
    & \textit{f32 - m7}                            & 33.69 & 33.53 & 29.60 & 29.52 & 28.81 & 28.76 & 27.16 & 27.10 \\
    & \textit{f32 - m11}                           & 33.81 & 33.63 & 29.68 & 29.60 & 28.86 & 28.80 & 27.35 & 27.27 \\
    & \textit{f64 - m11} (large)                   & 34.14 & 34.01 & 29.88 & 29.82 & 29.02 & 28.98 & 27.81 & 27.76 \\
    \hline
    \multirow{8}{*}{4$\times$} & \textit{f32 - m1} & 30.48 & 30.42 & 27.31 & 27.27 & 26.94 & 26.91 & 24.47 & 24.45 \\
    & \textit{f32 - m2} (small)                    & 30.84 & 30.82 & 27.55 & 27.54 & 27.07 & 27.06 & 24.74 & 24.74 \\
    & \textit{f32 - m3}                            & 31.04 & 30.95 & 27.65 & 27.58 & 27.16 & 27.12 & 24.90 & 24.86 \\
    & \textit{f32 - m5} (medium)                   & 31.27 & 31.21 & 27.79 & 27.76 & 27.24 & 27.21 & 25.08 & 25.06 \\
    & \textit{f32 - m7}                            & 31.39 & 31.29 & 27.83 & 27.80 & 27.30 & 27.27 & 25.22 & 25.18 \\
    & \textit{f32 - m11}                           & 31.50 & 31.36 & 27.93 & 27.85 & 27.35 & 27.29 & 25.32 & 25.27 \\
    & \textit{f64 - m11} (large)                   & 31.77 & 31.73 & 28.15 & 28.12 & 27.50 & 27.48 & 25.74 & 25.72 \\
    \hline
  \end{tabular}
    \caption{QuickSRNet PSNRs (dB) evaluated for different scaling factors ($2\times$, $3\times$, and $4\times$) on benchmark SISR datasets before and after quantization}
  \label{table:qsrnet-2x-3x-4x-psnr-fp16-int8}
\end{table*}

\begin{table*}[t]
  \centering
  \setlength{\tabcolsep}{1.5em}
  \footnotesize
  \begin{tabular}{| c | l | c  c  c  c |}
    \hline
    \textbf{Scaling} & \multirow{2}{*}{\textbf{QuickSRNet Specification}} & \textbf{Set5} & \textbf{Set14} & \textbf{BSD100} & \textbf{Urban100} \\
    \textbf{Factor} &  & \textbf{PSNR / SSIM} & \textbf{PSNR / SSIM} & \textbf{PSNR / SSIM} & \textbf{PSNR / SSIM} \\
    \hline
    \multirow{8}{*}{2$\times$} & \textit{f32 - m1} & 36.83 / 0.9563 & 32.35 / 0.9085 & 31.43 / 0.8900 & 29.66 / 0.8999 \\
    & \textit{f32 - m2} (small)                    & 37.12 / 0.9575 & 32.57 / 0.9107 & 31.61 / 0.8925 & 30.15 / 0.9067 \\
    & \textit{f32 - m3}                            & 37.30 / 0.9583 & 32.72 / 0.9117 & 31.72 / 0.8942 & 30.43 / 0.9104 \\
    & \textit{f32 - m5} (medium)                   & 37.39 / 0.9586 & 32.82 / 0.9130 & 31.82 / 0.8955 & 30.75 / 0.9142 \\
    & \textit{f32 - m7}                            & 37.51 / 0.9593 & 32.95 / 0.9136 & 31.88 / 0.8964 & 30.93 / 0.9164 \\
    & \textit{f32 - m11}                           & 37.59 / 0.9594 & 33.00 / 0.9142 & 31.95 / 0.8973 & 31.14 / 0.9186 \\
    & \textit{f64 - m11} (large)                   & 37.87 / 0.9603 & 33.29 / 0.9166 & 32.12 / 0.8992 & 31.74 / 0.9248 \\
    \hline
    \multirow{8}{*}{3$\times$} & \textit{f32 - m1} & 32.75 / 0.9112 & 29.08 / 0.8234 & 28.41 / 0.7880 & 26.19 / 0.8026 \\
    & \textit{f32 - m2} (small)                    & 33.10 / 0.9157 & 29.29 / 0.8282 & 28.57 / 0.7919 & 26.53 / 0.8128 \\
    & \textit{f32 - m3}                            & 33.33 / 0.9180 & 29.39 / 0.8298 & 28.67 / 0.7947 & 26.77 / 0.8195 \\
    & \textit{f32 - m5} (medium)                   & 33.58 / 0.9206 & 29.49 / 0.8327 & 28.75 / 0.7971 & 27.02 / 0.8266 \\
    & \textit{f32 - m7}                            & 33.69 / 0.9216 & 29.60 / 0.8335 & 28.81 / 0.7986 & 27.16 / 0.8303 \\
    & \textit{f32 - m11}                           & 33.81 / 0.9226 & 29.68 / 0.8346 & 28.86 / 0.8000 & 27.35 / 0.8347 \\
    & \textit{f64 - m11} (large)                   & 34.14 / 0.9258 & 29.88 / 0.8397 & 29.02 / 0.8038 & 27.81 / 0.8459 \\
    \hline
    \multirow{8}{*}{4$\times$} & \textit{f32 - m1} & 30.48 / 0.8659 & 27.31 / 0.7559 & 26.94 / 0.7147 & 24.47 / 0.7262 \\
    & \textit{f32 - m2} (small)                    & 30.84 / 0.8741 & 27.55 / 0.7635 & 27.07 / 0.7196 & 24.74 / 0.7382 \\
    & \textit{f32 - m3}                            & 31.04 / 0.8773 & 27.65 / 0.7656 & 27.16 / 0.7226 & 24.90 / 0.7447 \\
    & \textit{f32 - m5} (medium)                   & 31.27 / 0.8821 & 27.79 / 0.7699 & 27.24 / 0.7253 & 25.08 / 0.7517 \\
    & \textit{f32 - m7}                            & 31.39 / 0.8838 & 27.83 / 0.7709 & 27.30 / 0.7275 & 25.22 / 0.7573 \\
    & \textit{f32 - m11}                           & 31.50 / 0.8856 & 27.93 / 0.7729 & 27.35 / 0.7289 & 25.32 / 0.7619 \\
    & \textit{f64 - m11} (large)                   & 31.77 / 0.8908 & 28.15 / 0.7797 & 27.50 / 0.7344 & 25.74 / 0.7761 \\
    \hline    
  \end{tabular}
    \caption{QuickSRNet PSNRs (dB) and SSIM numbers evaluated for different scaling factors ($2\times$, $3\times$, and $4\times$) on benchmark SISR datasets before quantization}
  \label{table:qsrnet-2x-3x-4x-psnr-ssim-fp16}
\end{table*}

\newpage

\phantom{}

\newpage

\phantom{}

\newpage

\subsection*{Exporting QuickSRNet to ONNX for on-device profiling}

Before running the model on device, we shuffle the weights of some of the convolutional layers, before depth-to-space and after space-to-depth (for $1.5\times$ model) operations. This is necessary because the data layout of PyTorch’s depth-to-space operation (CRD) is not optimized on our target device (Hexagon Processor of a mobile device with Snapdragon 8 Gen 1). For better on-device performance, the data layout needs to be changed to DCR. The appropriate method of creating a QuickSRNet model instance with the shuffled weights (in DCR format) can be done with the following steps. Below are a bunch of prerequisites to accomplish this task:

\begin{itemize}
    \item The PyTorch implementation of QuickSRNet can be found \url{https://github.com/quic/aimet-model-zoo/blob/torch_transformer_quicksrnet/zoo_torch/examples/superres/utils/models.py\#L332-L365}
    \item Pre-trained weights (including AIMET-quantized weights and encodings) are available \url{https://github.com/quic/aimet-model-zoo/releases/tag/quicksrnet-checkpoint-pytorch}
    \item A Jupyter Notebook that shows how to load and use QuickSRNet is also available \url{https://github.com/quic/aimet-model-zoo/blob/torch_transformer_quicksrnet/zoo_torch/examples/superres/notebooks/superres_quanteval.ipynb}
\end{itemize}

\paragraph{Step 1} Load the quantized QuickSRNet model from the checkpointed weights and encodings. With the PyTorch implementation of QuickSRNet, the model can be instantiated with the appropriately shuffled weights as follows:
    
\begin{minted}[
baselinestretch=1.2,
fontsize=\scriptsize,
autogobble
]{python}
    import torch
    
    # Use one of QuickSRNetSmall, QuickSRNetMedium or QuickSRNetLarge with the desired scaling factor.
    scaling_factor = 2
    model = QuickSRNetSmall(scaling_factor=scaling_factor)
    
    state_dict = torch.load(model_checkpoint_path, map_location='cpu')['state_dict']
    model.load_state_dict(state_dict)
    model.to(device)    # `device` is one of 'cuda' or 'cpu'

    # Re-arrange the weights of the appropriate conv layer(s)
    model.to_dcr()

\end{minted}

\paragraph{Step 2 (\textit{optional})} To use QuickSRNet quantized using AIMET, use the following steps:

\begin{minted}[
baselinestretch=1.2,
fontsize=\scriptsize,
autogobble
]{python}
    dummy_input_shape = (1, 3, 256, 256)    # Expected input shape for the model (1 x C x H x W)
    dummy_input = torch.randn(dummy_input_shape)

    sim = QuantizationSimModel(model=model,
                               dummy_input=dummy_input,
                               quant_scheme=QuantScheme.post_training_tf_enhanced,
                               default_output_bw=8,
                               default_param_bw=8)
    sim.set_and_freeze_param_encodings(encoding_path=encoding_path)
    sim.compute_encodings(forward_pass_callback=pass_calibration_data,
                          forward_pass_callback_args=(calibration_data,
                                                      scaling_factor,
                                                      use_cuda))
    
\end{minted}
    
\paragraph{Step 3} Export the model to ONNX:

\begin{minted}[
baselinestretch=1.2,
fontsize=\scriptsize,
autogobble
]{python}
    import os
    import torch
    from aimet_torch.onnx_utils import OnnxExportApiArgs

    filename = "<onnx_filename>"
    output_dir = "<output_dir>"
    model_save_path = "<output_dir>/<filename>.onnx"

    # PixelUnshuffle does not map to space-to-depth without the code below
    import torch.onnx.symbolic_helper as sym_help
    import torch.onnx.symbolic_opset11 as opset11
    from torch.onnx.symbolic_helper import parse_args, _unimplemented
    
    @parse_args('v', 'i')
    def pixel_unshuffle(g, self, downscale_factor):
        rank = sym_help._get_tensor_rank(self)
        if rank is not None and rank != 4:
            return _unimplemented("pixel_unshuffle", "only support 4d input")
        return g.op("SpaceToDepth", self, blocksize_i=downscale_factor)
    opset11.pixel_unshuffle = pixel_unshuffle

    # Set `use_quantized` to `True` if exporting the quantized model, else `False`
    if use_quantized:
        sim.export(output_dir,
                   filename,
                   dummy_input,
                   onnx_export_args=OnnxExportApiArgs(opset_version=11))
    else:
        torch.onnx.export(model, dummy_input, model_save_path, export_params=True, opset_version=11)

\end{minted}

\paragraph{Step 4} Convert the ONNX space-to-depth and/or depth-to-space operations to DCR:

\begin{minted}[
baselinestretch=1.2,
fontsize=\scriptsize,
autogobble
]{python}
    import onnx
    from onnx.helper import make_attribute

    def overwrite_onnx_d2s_mode_to_dcr(onnx_path):
        """Manual override of the depth-to-space mode to DCR."""

        onnx_model = onnx.load(onnx_path)
        graph = onnx_model.graph
        for node in graph.node:
            if node.op_type == 'DepthToSpace':
                depth_to_space_attribute = node.attribute
    
                found = False
                for idx, attr in enumerate(node.attribute):
                    if attr.name == 'mode':
                        found = True
                        break
                if found:
                    node.attribute.pop(idx)
    
                new_attr = make_attribute('s', 'DCR')
                new_attr.name = 'mode'
    
                depth_to_space_attribute.extend([new_attr])
        onnx.save(onnx_model, onnx_path)

    onnx_path = "<output_dir>/<filename>.onnx"    # Path to the exported ONNX file
    overwrite_onnx_d2s_mode_to_dcr(onnx_path)

\end{minted}

\paragraph{Step 5} Re-order per-channel encodings for the quantized model to DCR:

\begin{minted}[
baselinestretch=1.2,
fontsize=\scriptsize,
autogobble
]{python}
import json

def reorder_per_channel_encodings_to_dcr(encodings_path, layer_names):
    """
    Used to re-arrange the per-channel encodings of the conv layer(s) preceding the final depth-to-space operation.
    
    This is necessary because the data layout of PyTorch’s depth-to-space operation (CRD) is 
    not optimized on device. For better on-device performance, the data layout needs to be changed 
    to DCR.
    
    Note: in the case of per-layer quantization, this function does not do anything.
    """
    
    with open(encodings_path) as f:
        encodings = json.load(f)
    
    new_encodings = encodings.copy()
    to_shuffle = [key for layer_name in layer_names for key in encodings['param_encodings'] if layer_name in key]
    for key in to_shuffle:
        per_channel_enc = encodings['param_encodings'][key]
        if len(per_channel_enc) > 1:
            scaling_factor = int((len(per_channel_enc) / 3) ** 0.5)
            new_encodings['param_encodings'][key] = [per_channel_enc[i + k * (scaling_factor ** 2)] 
                                                     for i in range(scaling_factor ** 2) for k in range(3)]
        else:
            # per-layer quantization: do nothing
            pass
    
    with open(encodings_path, 'w') as f:
        json.dump(new_encodings, f, sort_keys=True, indent=4)

reorder_per_channel_encodings_to_dcr(encodings_path, ['anchor', 'conv_last'])

\end{minted}

\end{document}